\documentclass[12pt]{iopart}

\usepackage{iopams}
\usepackage{graphicx}  
\begin{document}

\title[Graphene to Graphane: A Theoretical Study]{Graphene to Graphane: A Theoretical Study}

\author{M Z S Flores$^{1}$, P A S Autreto$^{1}$, S B Legoas$^{2}$ and D S Galvao$^{1}$}

\address{$^{1}$Instituto de F\'{i}sica ``Gleb Wataghin'', Universidade Estadual de Campinas, Unicamp, C.P. 6165, 13083-970, Campinas, S\~{a}o Paulo, Brazil}

\address{$^{2}$Centro de Ci\^{e}ncias e Tecnologia, Universidade Federal de Roraima, 69304-000, Boa Vista, Roraima, Brazil}

\ead{galvao@ifi.unicamp.br}

\begin{abstract}
Graphane is a two-dimensional system consisting of a single layer of fully saturated (sp$^3$ hybridization) carbon atoms. In an ideal graphane structure C-H bonds exhibit an alternating pattern (up and down with relation to the plane defined by the carbon atoms). In this work we have investigated  using \textit{ab initio} and reactive molecular dynamics simulations the role of H frustration (breaking the H atoms up and down  alternating pattern) in graphane-like structures. Our results show that significant percentage of uncorrelated H frustrated domains are formed in the early stages of the hydrogenation process leading to membrane shrinkage and extensive membrane corrugations. These results also suggest that large domains of perfect graphane-like structures are unlikely to be formed, H frustrated domains are always present.

\end{abstract}

\submitto{\NT}
\maketitle

\section{Introduction}

Carbon-based materials have been intensely investigated in recent decades. However, in spite of the enormous amount of theoretical and experimental works, the discovery of new structures seems endless, colossal carbon tubes \cite{colossal} and graphene \cite{geim} being recent examples.

Graphene is one of the most important subject in materials science today \cite{geim,geim1,castroneto}. It is a two-dimensional structure of sp$^2$ carbon atoms with very unusual and interesting electronic and mechanical properties.

It has been theoretically predicted that a related structure, called graphane \cite{sofo}, could exist in a stable form. Graphane consists of a single layer structure with fully saturated (sp$^3$ hybridization) carbon atoms with C-H bonds in an alternating pattern (up and down with relation to the plane defined by the carbon atoms). Its two most stable conformations are the so-called chair-like (H atoms alternating on both sides of the plane) and boat-like (H atoms alternating in pairs) \cite{sofo} (figure \ref{models}). A third member of these two-dimensional planar carbon structures, called graphyne \cite{ray1,ray2,ray3}, has also been predicted to exist but up to now only molecular fragments have been synthesized \cite{ray2}.

Indirect experimental evidences of graphane-like structures have been reported \cite{humberto,graphane2}. More recently, in a series of very elegant experiments, Elias \textit{et al.} \cite{geim2} demonstrated the existence of graphane formation from graphene membrane through its hydrogenation. They also demonstrated that this process is reversible. These fundamental discoveries open new and important perspectives to the use of graphene-based devices since the electronic gap values in graphanes could be controlled by the degree of hydrogenation \cite{geim2,comment}.

The Elias \textit{et al.} experiments consisted in exposing graphene membranes to H$^{+}$ from cold plasma. The H incorporation into the membranes results in altering the C sp$^2$ hybridizations to sp$^3$ ones. The experiments were also done with the membranes over SiO$_2$ substrates (only one membrane side exposed to H$^{+}$) and produced a material with different properties. Detailed studies of hydrogen atoms on graphene have been recently reported \cite{hyd1,hyd2,hyd3,hyd4,hyd5,hyd6,hyd7,hyd8}. 

This paper is organized as follows: in Sec. 2 we describe the used methodology. The results and discussions are presented in Sec. 3, followed by the summary and our conclusions in Sec. 4.

\section{Methodology}

We have used different methods to carry out our investigations.
Initially, we performed \textit{ab initio} quantum calculations in order to optimize the geometry of graphane-like structures. For comparison purposes, graphene structures were also calculated. Secondly, we used classical reactive bond-order approach in order to investigate the effects of hydrogenation on geometrical structures for a number of graphene membrane models.  Finally, molecular dynamics (MD) simulations were used to address the dynamics of hydrogen incorporation into graphene membranes.

We have carried out \textit{ab initio} total energy calculations in the framework of the density functional theory (DFT), as implemented in the DMol$^3$ code \cite{dmol3}. Exchange and correlation terms were treated within the generalized gradient (GGA) functional by Perdew, Burke, and Ernzerhof \cite{pbe}. Core electrons were treated in a non-relativistic all electron implementation of the potential. A double numerical quality  basis set with polarization function (DNP) were considered, with a real space cutoff of 3.7 \AA. The tolerances of energy, gradient, and displacement convergence were 0.00027 eV, 0.054 eV/\AA\ and 0.005 \AA, respectively.

We investigated fully and partially hydrogenated infinite (periodic boundary condition - PBC)  graphene structures, which requires the use of slab supercells. Sofo, Chaudhari, and Baker \cite{sofo} in their graphane work considered compact (interacting layers) structures. Here, in order to mimic the experimental conditions \cite{geim2}, we have considered isolated (non-interacting) layers.
For all cases considered here, the \textit{c} axis was fixed at 20 \AA (large enough to prevent interactions among different layers), and the remaining free parameters were fully optimized (figure \ref{models}). Internal atomic positions were free to vary in all the geometry minimization calculations. The total energy results as function of the unit cell volumes were fitted following the well known Murnaghan procedure \cite{murnaghan}. We have also considered small finite structures (figure \ref{fragment}) with hydrogen passivated borders. In order to investigate larger structures, where \textit{ab initio} quantum calculations becomes computationally prohibitive, we used ReaxFF binding energy bond order (BEBO) method \cite{duin1,duin2,duin3}.

ReaxFF is similar to standard non-reactive force fields, like MM3 \cite{mm3}, where the system energy is divided into partial energy contributions associated with, amongst others, valence angle bending, bond stretching, as well as, non-bonded van der Waals and Coulomb interactions \cite{duin1,duin2}. However, one main difference is that ReaxFF can handle bond formation and dissociation (making/breaking bonds) as a function of bond order values. ReaxFF was parameterized against DFT calculations, the average deviation between the predicted ReaxFF  heat of formation values and the experimental ones are of 2.8 and 2.9 Kcal/mol for non-conjugated and conjugated systems, respectively \cite{duin2}. ReaxFF is a reactive force field developed by Adri van Duin, William Goddard III and co-workers for use in MD simulations. This method allows the simulation of many types of chemical reactions. Similarly to the procedures we adopted in DFT calculations we have considered both finite and infinite (PBC) structures. We have carried out geometry optimizations using gradient conjugated techniques (convergence condition, gradient values less than 10$^-$$^3$).

The dynamics of hydrogen incorporation on graphene layers was studied under different conditions of temperature (300K, 500K and 650K) and hydrogen atmospheres (number of H atoms up to twice the number of carbon ones). We considered H distributed over one and both sides of the graphene layers (infinite membranes with 11 $\times$ 11 unit cells). For each temperature different H velocity distributions were used. In order to speed up the hydrogen incorporation in the first 500 fs of the simulations, after each MD run of 10 fs, the hydrogen velocities were recalculated and the H atoms directed towards the graphene membrane. Typical total MD runs were of 50 ps, timesteps of 0.5 fs, and using a Berendsen thermostat \cite{berendsen}.

\section{Results and Discussions} 

\begin{figure}
\begin{center}
\includegraphics[width=10cm]{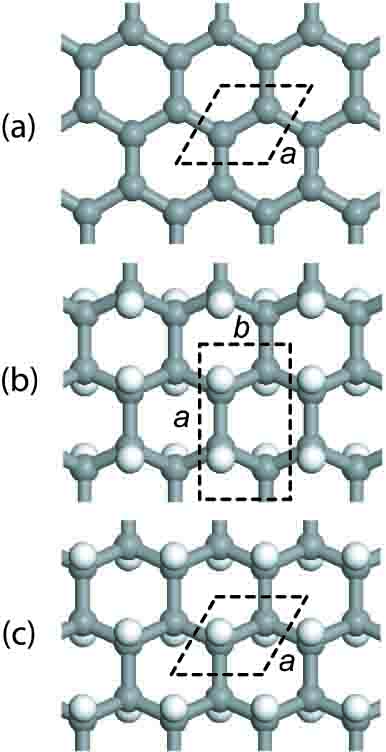}
\caption{Structural carbon membrane models considered in the DMol${^3}$
geometry optimization calculations. (a) Graphene, having two atoms per unit cell;
(b) graphane boat-like, with four carbon atoms and four hydrogen atoms per unit cell; (c) graphane chair-like, with four (two C and two H) atoms per unit cell. The dashed lines indicate the corresponding unit cell.
\textit{a} and \textit{b} refer to the lattice parameters. See text for discussions.}
\label{models}
\end{center}
\end{figure}

We started carrying out DMol$^3$ calculations for the infinite (PBC) structures shown in figure \ref{models}. 
The results are displayed in table \ref{1st_table}. The chair-like graphane (figure \ref{models}(c)) is more stable than the boat-like (figure \ref{models}(b)) by 0.03 Ha ($\sim 0.82$ eV). The G-boat presents two non equivalent carbon-carbon distances due to the existence of two different C-H alignments. Although the C-C distances in graphane are much longer (8 \%) than the ones present in graphene, its cell parameter is just slightly larger (3 \%) than the latter due to out-of-plane topology. We have also considered the case of the minimum unit cell with H atoms parallelly aligned (just one side of the membrane). Our results show that this system is unstable with the tendency of H$_2$ recombination and/or C-C breaking bonds. The results are in good agreement with previous work \cite{sofo,hyd1,hyd2,gra} and with the available experimental data \cite{geim,geim2}.

\begin{figure}
\begin{center}
\includegraphics[width=10cm]{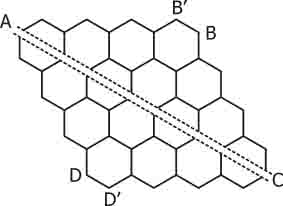}
\caption{Schematic draw of a graphene fragment before hydrogenation. The letters are the reference points for the distances displayed in tables \ref{2nd_table} and \ref{Pedro_table}. The diagonal bi-line (joining A and C points) is just to indicate that
the size of the membranes are different depending of the calculation being considered (DMol$^3$ or ReaxFF).}
\label{fragment}
\end{center}
\end{figure}

One important aspect to be investigated is how different hydrogenation patterns affect the geometry of the graphene membranes. One practical way to do this is to measure some representative distances, as the ones indicated in figure \ref{fragment} and the second neighbor C-C distances (which would correspond to the lattice parameter in ideal crystalline structures). 

\begin{figure}
\begin{center}
\includegraphics[width=10cm]{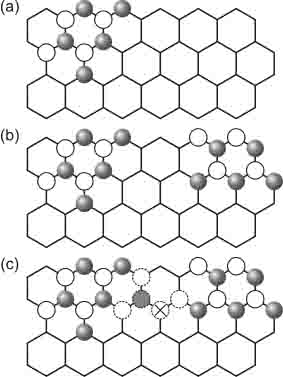}
\caption{Scheme of the formation of H frustrated domains. Closed circles refer to up hydrogen atoms, and open circles refer to down ones, with relation to the plane defined by the carbon atoms. Carbon atoms are omitted for clarity. (a) Initial stage of the hydrogen incorporation. (b) Hydrogenation
occurring at different regions. (c) H frustrated site, shown by an open triangle. See text for discussions.}
\label{frustration}
\end{center}
\end{figure}

\begin{figure}
\begin{center}
\includegraphics[width=10cm]{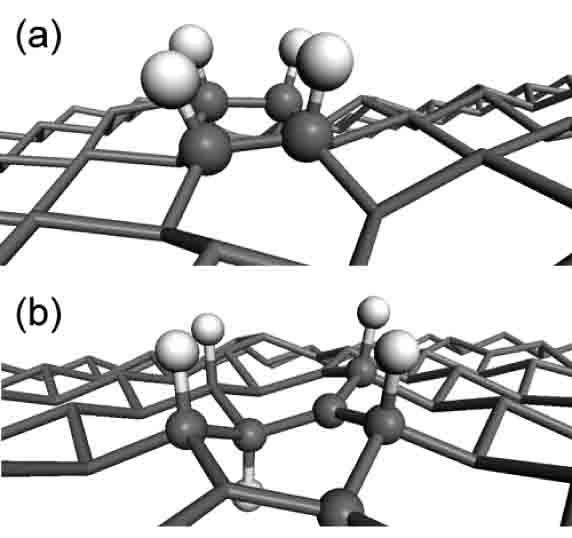}
\caption{Examples of different possible frustration types. (a) Frust-1, H frustration with parallel first-neighbor H atoms; (b) Frust-0, H frustration with `missing' first-neighbor H atoms. Atoms in the defect region are shown in ball and stick rendering. For view clarity the H atoms outside this region were made transparent.} 
\label{frustration2}
\end{center}
\end{figure}

If we consider that H atoms are randomly incorporated during plasma exposure \cite{geim}, there is a significant probability for the existence of H frustration (figure \ref{frustration}), which is a configuration where the sequence of alternating up and down H atoms is broken (frustrated) (see movie01 \cite{material}). This is similar to spin frustration in magnetic materials \cite{spin}. Two different H frustration configurations are possible (figure \ref{frustration2}), one with parallel H atoms (Frust-1) and the other with missing H atoms (Frust-0).

In figure \ref{frustration}(a) we show a domain of up and down H atoms. It is expected that after the first (up or down) H incorporation, the next favorable site is its first inverse neighbor (down or up), and so on. If the system is large enough uncorrelated domains might be formed (figure \ref{frustration}(b)). As the H coverage is continued it could occur that it is no longer possible the alternating sequence of up and down H atoms (figure \ref{frustration}(c)).

We have investigated finite fragments with and without H frustrations. We analyzed the associated geometrical changes (figure \ref{fragment}) in order to determine whether the structures expand or contract with relation to an equivalent ideal graphane structure. 
We have carried out DMol$^3$ (table \ref{2nd_table}) and ReaxFF (for larger fragments) calculations (table \ref{Pedro_table}, figure \ref{fragment}).
As we can see from these tables, DMol$^3$ and ReaxFF show similar and consistent results. The H frustration increases out of plane distortions which induce in-plane geometrical shrinkage. This effect is amplified when first-neighbor H atoms are parallelly aligned (figure \ref{frustration2}). 

\begin{figure}
\begin{center}
\includegraphics[width=10cm]{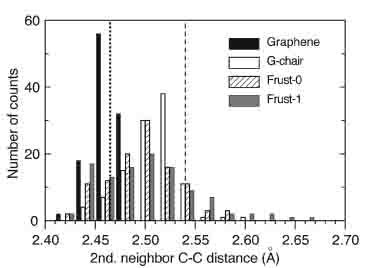}
\caption{Distribution of the second neighbor carbon-carbon distances of the
finite fragments listed in table \ref{2nd_table}. Vertical lines indicate the
lattice parameter values of ideal infinite graphene (dotted) and chair-like graphane (dashed), respectively.}
\label{histDMol}
\end{center}
\end{figure}

The representative distances indicated in (figure \ref{fragment}) (table \ref{2nd_table} and table \ref{Pedro_table}) provide a general view of the geometric changes produced by the different hydrogenation patterns. A more local information can be obtained averaging the C-C second-neighbor distances (which would correspond to the lattice parameter for ideal structures). As can be seen from the tables and from (figure \ref{histDMol}) the H frustration systematically reduce these distances to smaller values in comparison to ideal graphane structures. For particular configurations these distances can be even smaller than the corresponding graphene values ((see also supplementary materials \cite{material}).

\begin{figure}
\begin{center}
\includegraphics[width=10cm]{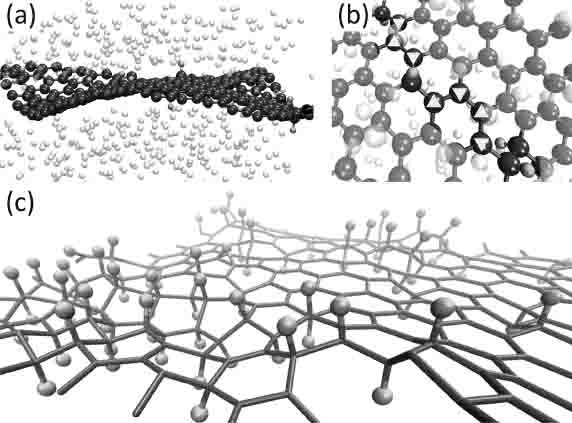}
\caption{(a) Representative snapshot of the early hydrogenation stages from ReaxFF molecular simulations at 500 K. Non-bonded atomic H atoms are indicated in white and C-bonded ones in green. (b) Zoomed region indicating H frustrated domains formed. The triangle path shows that a sequence of up and down H atoms is no longer possible. (c) Representative snapshot of the final hydrogenation states. Extensive hydrogenation and multiple formed H domains are clearly visible.}
\label{MD}
\end{center}
\end{figure}

Finally, we investigated the dynamics of H incorporation using MD simulations.
We have carried out MD calculations to investigate the formation of graphane-like through hydrogen reactions with the C carbons of the graphene layers. We have used infinite (using periodic boundary conditions) graphene structures. 
In figure \ref{MD} we show representative snapshots from the early and final stages of a simulation at 500 K (see movie02 \cite{material}). The results show that significant percentage of uncorrelated H frustrated domains are formed in the early stages of the hydrogenation processes leading to membrane shrinkages and extensive membrane corrugations. These results also suggest that large domains of perfect graphane-like structures are unlikely to be formed, H frustrated domains are always present. The number of these domains seems to be sensitive to small variations of temperatures and H gas densities. We run annealing cycle simulations to analyze the stability of these domains once formed. Our results show that H frustrated domain are very stable, high temperatures are needed to reverse (dehydrogenation processes) graphane-like structures to their original graphene configurations. 

\section{Summary and Conclusions}

We have performed geometry optimizations and molecular dynamics simulations using ab initio DMol$^3$ and classical reactive bond order ReaxFF, respectively, for the hydrogenation process of graphene leading to graphane-like structures. Graphane is a two-dimensional system consisting of a single planar layer of fully saturated (sp$^3$ hybridization) carbon atoms with H atoms attached to them in an alternating pattern (up and down with relation to the plane defined by the carbon atoms).

Our results show that H frustration are very likely to occur. H frustration is a configuration where the sequence of alternating up and down H atoms is broken (frustrated). The H frustration increase out of plane distortions (in relation to ideal graphane structures) which induces in-plane dimensional shrinkage. The net result is a decrease of the carbon-carbon distances in relation to the ideal graphane values. This effect is amplified when first neighbor H atoms are parallelly aligned. The results show that significant percentage of uncorrelated H frustrated domains are easily formed in the early stages of the hydrogenation process leading to lattice decreased values and extensive membrane corrugations. These results also suggest that large domains of perfect graphane-like structures are unlikely to be formed, H frustrated domains are always present. The molecular dynamics simulations of the hydrogenation showed that one formed hydrogenated domains are very stable.

\section*{Acknowledgments}
This work was supported in part by the Brazilian Agencies CNPq, CAPES and FAPESP. The authors wish to thank Prof. A. van Duin for his very helpful assistance with the ReaxFF code, PASA and MZSF thanks his kind hospitality in Pennsylvania. DSG wish also to thank Prof. A. Geim for helpful discussions.

\newpage

\begin{table}[h]
\centering
\caption{DMol$^3$ results for the crystalline structures shown in figure \ref{models}. The energy per atom in the unit cell, the cell parameter values and the carbon-carbon distances are displayed. G-chair and G-boat refer to chair-like and boat-like graphane systems, respectively. More detailed geometrical data are provided in the supplementary materials \cite{material}.}
\begin{indented}
\item[]\begin{tabular}{llll}
%\hline
\br
    & \multicolumn{1}{c}{Graphene} & \multicolumn{1}{c}{G-chair} & \multicolumn{1}{c}{G-boat} \\ 
%\hline
\mr
Energy (Ha)  & \multicolumn{1}{c}{-304.68} & \multicolumn{1}{c}{-309.41} & \multicolumn{1}{c}{-309.38} \\ 
Lattice parameters: & & & \\
\multicolumn{1}{c}{\textit{a} (\AA)}  & \multicolumn{1}{c}{2.465} & \multicolumn{1}{c}{2.540} & \multicolumn{1}{c}{4.346} \\ 
\multicolumn{1}{c}{\textit{b} (\AA)}  & \multicolumn{1}{c}{2.465} & \multicolumn{1}{c}{2.540} & \multicolumn{1}{c}{2.509} \\ 
\multicolumn{1}{c}{$\gamma$ ($^\circ$)} & \multicolumn{1}{c}{120} &
\multicolumn{1}{c}{120} & \multicolumn{1}{c}{90} \\
 C-C bond length (\AA)      & \multicolumn{1}{c}{1.423} & \multicolumn{1}{c}{1.537} & \multicolumn{1}{c}{\textrm{1.581, 1.537}} \\
%\hline
\br
\end{tabular}
\label{1st_table}
\end{indented}
\end{table}

\newpage

\begin{table}[h]
\centering
\caption{Distances (in \AA) between reference points for the molecular systems depicted in figure \ref{fragment} calculated with DMol$^3$. 
G-chair (39.6\% H) and G-boat (37.5\% H) are related to the structures in table \ref{1st_table}.
Frust-1 (22.9\% H) and Frust-0 (20.8\% H) refer to parallel (figure \ref{frustration2}(a)) and missing (figure \ref{frustration}(b)) hydrogen atoms in frustrated domains, respectively. 
Parameters $\bar{\textrm{d}}$ and $\bar{\textrm{a}}$ (in \AA) are the mean value of the 1st. and 2nd. neighbor carbon-carbon distances, respectively. Values in parentheses refer to the standard deviation.
More detailed geometrical data are provided in the supplementary materials \cite{material}}
\begin{indented}
\item[]\begin{tabular}{llllllll}
%\hline
\br
System & \multicolumn{1}{c}{d$_{A-B}$} & \multicolumn{1}{c}{d$_{B'-C}$} & \multicolumn{1}{c}{d$_{C-D}$} & \multicolumn{1}{c}{d$_{D'-A}$} & \multicolumn{1}{c}{d$_{B-D}$} & \multicolumn{1}{c}{$\bar{\textrm{d}}$} &
\multicolumn{1}{c}{$\bar{\textrm{a}}$} \\
%\hline
\mr
Graphene & \multicolumn{1}{c}{9.804} & \multicolumn{1}{c}{9.799} & \multicolumn{1}{c}{9.804} & \multicolumn{1}{c}{9.799} & \multicolumn{1}{c}{9.974} & 1.417(19) & 2.462(12) \\ 
G-chair & \multicolumn{1}{c}{9.861} & \multicolumn{1}{c}{9.841} & \multicolumn{1}{c}{9.882} & \multicolumn{1}{c}{9.847} & \multicolumn{1}{c}{10.050} & 1.470(61) & 2.505(27) \\ 
G-boat & \multicolumn{1}{c}{9.852} & \multicolumn{1}{c}{9.818} & \multicolumn{1}{c}{9.852} & \multicolumn{1}{c}{9.818} & \multicolumn{1}{c}{9.977} & 1.470(65) & 2.513(43) \\ 
Frust-0 & \multicolumn{1}{c}{9.788} & \multicolumn{1}{c}{9.857} & \multicolumn{1}{c}{9.876} & \multicolumn{1}{c}{9.823} & \multicolumn{1}{c}{10.003} & 1.448(54) & 2.496(35) \\ 
Frust-1 & \multicolumn{1}{c}{9.740} & \multicolumn{1}{c}{9.802} & \multicolumn{1}{c}{9.866} & \multicolumn{1}{c}{9.786} & \multicolumn{1}{c}{9.990} & 1.451(59) & 2.500(48) \\ 
%\hline
\br
\end{tabular}
\label{2nd_table}
\end{indented}
\end{table}

\newpage

\begin{table}[h]
\centering
\caption{Distances (in \AA) between the reference points for the system depicted in figure \ref{fragment} calculated with ReaxFF. G-chair refers to the chairlike graphane system. Frust-0 and Frust-1 refer to defects-like shown in (figure \ref{frustration2}). The number in parenthesis indicate the number of frustrated domains in the structure. Frust-1-in-line and Frust-0-in-line refer to frustation (of type 1 and 0, respectively) created along a line through the graphene membrane.
The Graphane-in-Graphene refers to small region of graphane structure into a graphene membrane. 
More detailed geometrical data are provided in the supplementary materials \cite{material}.}
\begin{indented}
\item[]\begin{tabular}{llllllll}
%\hline
\br
System & \multicolumn{1}{c}{d$_{A-B}$} & \multicolumn{1}{c}{d$_{B'-C}$} & \multicolumn{1}{c}{d$_{C-D}$} & \multicolumn{1}{c}{d$_{D'-A}$} & \multicolumn{1}{c}{d$_{A-C}$} & \multicolumn{1}{c}{d$_{B-D}$} &
\multicolumn{1}{c}{$\bar{\textrm{a}}$} \\ 
%\hline
\mr
Graphene & \multicolumn{1}{c}{22.26} & \multicolumn{1}{c}{22.26} & \multicolumn{1}{c}{22.26} & \multicolumn{1}{c}{22.26} & \multicolumn{1}{c}{37.08} & \multicolumn{1}{c}{22.47}  \\ 
G-chair & \multicolumn{1}{c}{22.95} & \multicolumn{1}{c}{22.94} & \multicolumn{1}{c}{22.95} & \multicolumn{1}{c}{22.94} & \multicolumn{1}{c}{38.40} & \multicolumn{1}{c}{23.01} \\ 
Frust-1 (13) & \multicolumn{1}{c}{22.21} & \multicolumn{1}{c}{22.92} & \multicolumn{1}{c}{22.80} & \multicolumn{1}{c}{22.90} & \multicolumn{1}{c}{37.95} & \multicolumn{1}{c}{22.47}  \\ 
Frust-0 (13) & \multicolumn{1}{c}{22.38} & \multicolumn{1}{c}{22.98} & \multicolumn{1}{c}{22.77} & \multicolumn{1}{c}{22.82} & \multicolumn{1}{c}{38.12} & \multicolumn{1}{c}{22.37}  \\
Frust-1-in-line & \multicolumn{1}{c}{22.82} & \multicolumn{1}{c}{20.93} & \multicolumn{1}{c}{22.98} & \multicolumn{1}{c}{22.92} & \multicolumn{1}{c}{37.84} & \multicolumn{1}{c}{22.81}  \\ 
Frust-0 in line & \multicolumn{1}{c}{23.00} & \multicolumn{1}{c}{21.22} & \multicolumn{1}{c}{23.41} & \multicolumn{1}{c}{22.98} & \multicolumn{1}{c}{35.12} & \multicolumn{1}{c}{22.34}  \\ 
Graphane-in-Graphene & \multicolumn{1}{c}{22.15} & \multicolumn{1}{c}{22.07} & \multicolumn{1}{c}{21.93} & \multicolumn{1}{c}{22.07} & \multicolumn{1}{c}{36.58} & \multicolumn{1}{c}{22.26}  \\ 
%\hline
\br
\end{tabular}
\label{Pedro_table}
\end{indented}
\end{table}

\newpage

\clearpage

\newpage

\clearpage

\clearpage

\clearpage

\end{document}